
\input Jnl.TEX
\input Reforder.TEX

\def\dd{d^{\dagger}_{i\sigma}}

\def\d{d_{i\sigma}}

\def\pd{p^{\dagger}_{j\sigma}}
\def\pdp{p^{\dagger}_{j'\sigma'}}
\def\p{p_{j\sigma}}

\def\ddm{d^{\dagger}_{i\bar{\sigma}}}
\def\dm{d_{i\bar{\sigma}}}
\def\pdm{p^{\dagger}_{j\bar{\sigma}}}
\def\pmm{p_{j\bar{\sigma}}}

\def\s{\sigma}
\def\D{\Delta}
\def\ed{- {\Delta\over 2}}
\def\ep{{\Delta\over 2}}
\def\inf{\infty}
\def\dag{\dagger}
\def\vd{ {V\over\D}}
\def\thd{{1 \over 3}}
\def\eps{\epsilon}

\title {  Strong Coupling Descriptions of High Temperature
Superconductors: Electronic Attraction from a repulsive potential.
}

\author W. Barford
\affil {\rm Department of Physics, University of Sheffield,
Sheffield, S3 7RH,
United Kingdom.}

\author M. W. Long
\affil {\rm School of Physics and Space Research, University of
Birmingham, Birmingham, B15 2TT, United Kingdom.}

\abstract{ }

\vskip 3.0 truein


October, 1992.

\endtopmatter

\head{Abstract}
We consider the effect of the nearest neighbour copper-oxygen
repulsion, $V$, when coupled to the charge transfer resonances
Cu$^{2+}$ $\to$ Cu$^{3+}$ and  Cu$^{2+}$ $\to$ Cu$^{+}$ in the high
temperature cuprate superconductors. This is done by deriving
effective low energy Hamiltonians correct to second order in the
copper-oxygen hybridisation. Only hole doping is considered. When
Cu$^{2+}$ $\to$ Cu$^{3+}$ fluctuations dominate we derive an
effective one-band model of `Zhang-Rice' singlets with a nearest
neighbour repulsion between these singlets. When Cu$^{2+}$ $\to$
Cu$^{+}$ fluctuations dominate we find rich and complex behaviour.
If $0 \ll \vd \leq {1 \over 2}$ (where $\Delta$ is the `bare'
copper-oxygen charge transfer gap) we show that clusters of charge
are more stable than isolated charges. On the other hand, if $0 \leq
\vd \ll {1 \over 2}$ the Hamiltonian contains both weak attractive
and repulsive two body potentials. Calculations on clusters indicate
that the attractive potentials have the same correlations as the more
dominate `single particle' terms suggesting the possibility of `s'
wave pairing.

\vfill\eject

\head {1. Introduction}

An unusual feature of the copper-oxide high temperature
superconductors is the close proximity of the atomic copper and
oxygen energy levels. This leads to charge transfer resonances of
the type Cu$^{2+}$ $\to$ Cu$^{3+}$ and  Cu$^{2+}$ $\to$ Cu$^{+}$.
Varma, Schmitt-Rink and Abrahams\refto{varma} were the first to
offer general arguments that these charge transfer resonances,
coupled to the nearest neighbour copper-oxygen repulsion, could
lead to `s-wave' superconductivity. Numerical\refto{hi, ba} and mean
field calculations\refto{little, grilli} confirmed that charge
clustering and superconducting instabilities are indeed a possibility
for certain parameter ranges.

In this paper we address this issue by deriving an effective low
energy Hamiltonian from the natural tight binding Hamiltonian for the
copper-oxide planes\refto{em} which includes the copper and
oxygen orbitals on an equal footing. This is:
$$
H_0 = \ed \sum_{i\s} \dd\d + U \sum_i \dd\ddm\dm\d + \ep \sum_{j\s}
\pd\p + U_p \sum_j \pd\pdm\pmm\p
$$
$$
+ V \sum_{<ij>\s\s'}\dd\d p^{\dagger}_{j\sigma'}p_{j\sigma'}
\eqno(1.1a)
$$
$$
H_1 = t \sum_{<ij>\s} (\dd\p + \pd\d),
\eqno(1.1b)
$$
where $\dd$ and $\pd$ create ${\it holes}$ in the relevant
hybridising orbitals on copper (site label $i$) and oxygen (site
label $j$) atoms respectively. The vacuum of (1.1) are the closed
shells $d^{10}$ (Cu$^{+}$) and $p^6$ (O$^{2-}$). $\Delta$ is the
`bare' charge transfer gap, $t$ is the hopping matrix element
connecting copper and oxygen sites, $V$ is the nearest neighbour
copper-oxygen coulomb repulsion and the  $U$s are the onsite coulomb
repulsions. Spectroscopic evidence\refto{sw} and fittings to
constrained density functional calculations\refto{mms} suggest that
$U \sim 9$ eV, $U_p \sim 5$ eV, $V \sim 0.8$ eV, $\D \sim 3$ eV and
$t \sim 1.5$eV.  Another term, which is sometimes considered, is the
direct oxygen-oxygen hopping matrix element, $t_{pp} \sim 0.6$ eV,
which has the effect of renormalising the charge transfer gap. This
term will not be considered in this paper. The `parent' compound
has one hole in every copper orbital ($d^9$) and empty oxygen
orbitals ($p^6$).

Our approach is to treat $t$ perturbatively, relegate
all the other energy scales to zero or infinity and perform a
Schrieffer-Woolf canonical transformation\refto{k, ck} on
(1.1) in the presence of the nearest neighbour coulomb repulsion.
This
 derives an
effective Hamiltonian whose leading order hopping term is O($t^2$),
and where there are residual many body effects resulting from the
copper-oxygen repulsion. The restriction to second order in the
hybridisation does mean, however,
that we cannot include the antiferromagnetic super-exchange effects
directly; we  therefore
only consider the effects caused by  the motion of charge
carriers. In principle, spin fluctuations can be included on an
{\it ad hoc} basis
 by adding a Heisenberg term to the effective Hamiltonians.

This method has been used by Fedro and Sch\"uttler\refto{fs} and
the present authors\refto{lb1} to derive an effective Hamiltonian for
electron doping. In this case the low energy physics is described
by a one band Hubbard Model with a nearest neighbour repulsion of
the charge carriers which reside on the copper sites. In this paper
we only consider hole doping. In $\S 2$ the motion of charge
carriers via virtual Cu$^{3+}$ excitations is considered. We find
that an essentially single band model describing the motion of the
`Zhang-Rice' singlet\refto{zr}, with a nearest neighbour repulsion
of these singlets, is appropriate. In $\S\S$ 3, 4 and 5 a
detailed discussion of the motion of charge carriers via virtual
Cu$^{+}$ excitations is given.  We find quite different and very rich
behaviour. First, even the motion of a single oxygen hole becomes
difficult to analyse, rendering the `Zhang-Rice' analysis
incomplete. Second,  when $0 \ll \vd \leq {1 \over 2}$ we find that
charge clustering occurs. Third, when $0 \leq \vd \ll {1 \over 2}$
we find evidence from cluster calculations for weak coupling `s-wave'
superconductivity, in agreement with the mean field
analyses\refto{little, grilli}. The conclusions of this work have
been presented elsewhere\refto{lb}. Finally, we conclude this paper
in $\S 6$.

\head{ 2. Motion  of Holes via Cu$^{3+}$ Excitations}

In the two band model (1.1) in the atomic limit
there are two natural charge states for the added hole. The first
possibility is that the copper sites dominates with the holes
forming Cu$^{3+}$. The second is that both copper and oxygen
sites must be considered with the added hole residing on oxygen
sites as O$^-$, and these holes move about in a spin background
residing on the copper sites. The second case will only be
considered as this is experimentally more justifiable\refto{sw}.
With the assumption that the holes reside predominately on oxygen
sites there are still two important limits to  consider as the
hybridisation is increased. If the direct oxygen-oxygen
hybridisation is small then delocalisation occurs across copper
sites. There are two distinct mechanisms for hole motion. The first
is to assume that added holes hop between oxygen sites via an
intermediate state in which the copper site is doubly occupied with
holes, namely Cu$^{3+}$. The second is to assume that in the
intermediate state a hole vacates the copper site leaving a closed
shell, namely Cu$^{+}$. The experimental evidence is in favour of
the latter, but in this section we consider the former possibility.
We treat only virtual occupancy of the intermediate charge states,
seperating out the two physical effects by allowing the copper
coulomb repulsion, $U$, to diverge. In this section we enforce this
by allowing $\D \rightarrow \inf$ while $(U-\D) < \inf$.

The assumptions that each new d-state is singly occupied and that
the intermediate state involves the Cu$^{3+}$ gives, by the second
order canonical transformation\refto{k, ck} (details may be found
in reference\refto{lb1}),
$$
H = - t^2 \sum_{<ij>\s}\sum_{<ij'>\s}
S^{\dagger}_{ij}[f^{-1}_{ij}+f^{-1}_{ij'}]S_{ij'}
\eqno(2.1a)
$$
for the $O(t^2)$ part of the Hamiltonian, with the `singlet' operator
being defined as, $$
S_{ij}^{\dagger} = {1\over \sqrt 2}\sum_\s {\rm sgn}(\s) \pd\ddm
\eqno(2.1b)
$$
and the matrix element,
$$
f_{ij} = U-\D -2V - U_pP_j + V \sum_{<ij>} P_l ,
\eqno(2.1c)
$$
where $P_j = \sum_\s \pd\p$ is the oxygen hole operator.

The form of this description may be readily understood. Since the
intermediate state involves two holes on a copper site the relevant
oxygen hole must be in a singlet configuration with respect to the
copper site that hybridises with it. The $S_{ij}^{\dagger}$
operators create precisely the relevant spin configuration.

In order to understand the matrix elements, however, we study the
states which the `new' operators create. The $\dd$ creates a pure
copper hole in this limit. The energy gap between this hole and the
neighbouring oxygen levels is assumed to be infinite, so the
hole gains nothing by hybridising. The $\pd$ creates a hole which is
predominately on the oxygen site but is optimally hybridised onto
the two neighbouring copper sites. When the `oxygen' hole resides
on an oxygen site it is repelled by both neighbouring copper holes.
However, when it resides on a copper site it is repelled by the
holes on neighbouring oxygen sites. The first consequence of this
repulsion is that doped holes only reside on the oxygen sites
provided that $U > \D +2V$, so the nearest neighbour repulsion
tends to push the doped holes onto the copper sites in this limit.
The second and more important consequence is that the nearest
neighbour repulsion between the copper and oxygen holes induces a
many body repulsion between charge carriers. This is because when
several `oxygen' holes neighbour the same copper site partial
occupation by one of those holes on the copper site results in a
repulsion between that hole and all the other `oxygen' holes. This
is a peculiarity to the topology of the copper oxide lattice: each
copper site has four neighbouring oxygen sites, but each oxygen
site has only two neighbouring copper sites.

We may derive the repulsion directly by perfoming the inversion of
the operator $f_{ij}$ (2.1c). Since $f_{ij}$ is a sum of commuting
operators the inversion is trivial, and a general prescription is
derived in the appendix. For the case that $U_p \rightarrow \inf$ we
find:
$$ H_1 = -D_1
\sum_{<ij>} \sum_{<ij'>} S_{ij}^{\dagger}S_{ij'}
\eqno(2.2a)
$$
$$
H_2 = D_2 \sum_{<ijl>} \sum_{<ij'l>} S_{ij}^{\dagger}P_l S_{ij'}
\eqno(2.2b)
$$
$$
H_3 = -D_3 \sum_{<ijlm>} \sum_{<ij'lm>} S_{ij}^{\dagger}P_l P_m
S_{ij'}
\eqno(2.2c)
$$
$$
H_4 = D_4 \sum_{<ijlmn>} S_{ij}^{\dagger}P_l P_m P_n S_{ij'}
\eqno(2.2d)
$$
where
$$
D_1 = {2t^2\over U-\D -2V},\hskip 0.5 truein
 D_{\rm n+1}= {V \over U
-\D +(n-2)V}D_n ,
\eqno(2.2e)
$$
from which it is evident that the two particle interactions
(represented in $H_2$) are repulsive.

Let us now consider the solution to the single particle problem
for this lattice. The basic conceptual difficulty is that the
oxygen hole wants to be simultaneously in a singlet configuration
with both of its neighbouring copper holes, and this is
impossible. A simple description can only be achieved by
considering states with local singlets (see figure 1). The problem
with this description is that it is non-orthogonal. The state in
figure 1(a) has probabilities 1 and 1/2 of being in a singlet with
the left and right copper holes. The state in figure 1(b) is
linearly independent with the probabilities reversed. These two
states form a complete basis but have a non-trivial overlap.

There are several excitations which are easy to describe and which
may be thought of as `non-bonding'. In figure 1(c) three phase
combinations of the four local singlet states connected to the
central copper atom are shown. Any state which corresponds to a
closed curve is an eigenstate of the Hamiltonian with zero
hybridisation energy. Several such states are depicted in figure
1(d). There is a natural seperation of these states into those where
the singlets are centred on the anti-ferromagnetic sublattice and
those centred on the other sublattice. These states take care of
three quarters of such states leaving the combination depicted in
figure 1(e). In this limit the states of Zhang and Rice\refto{zr}
can be justified, as also shown by Zhang\refto{z}. All we need now
is the Hamiltonian restricted to these states.

Denoting the uniform phase sum of figure (1e) as $S^{\dagger}_i$ (the
`Zhang-Rice' singlet), we find that at the one oxygen hole level:
$$
H= -8{\tilde t}\sum_i S^{\dagger}_iS_i - {\tilde t} \sum_{<ii'>\s}
S^{\dagger}_i d_{i'\s}\dd S_{i'},
\eqno(2.3a)
$$
$$
{\tilde t} = {t^2\over U -\D -2V},
\eqno(2.3b)
$$
where we ${\it assume}$ that the $S^{\dagger}_i$s are orthogonal.
The $S^{\dagger}_i$ may be considered as a `vacancy' in the one band
Hubbard model in the strong coupling limit. Indeed the spectrum is
identical since the orthogonality matrix is invertible and the
solutions to the problem in which the $S^{\dagger}_i$s are assumed
orthogonal also diagonalises the non-orthogonal problem. The
Nagaoka theorem\refto{n} therefore holds, and the `vacancy' spectrum
of the ferromagnetic branch is
$$
\epsilon_{\underline k} =
 -8{\tilde t} -2{\tilde t}[\cos (2k_x a) + \cos
(2k_ya)] + \ep
\eqno(2.4)
$$
with a ground state at $-12{\tilde t}+ \ep$  when
${\underline k}={\underline 0}$.

Although at the one particle level the spectrum is identical to
that of the Hubbard model, it is noteworthy that at the two particle
level the non-orthogonality becomes important. This results in a
nearest neighbour repulsion between charge carriers, in addition to
that resulting from the copper-oxygen repulsion.

\head {3. Motion of Holes via Cu$^+$ Excitations: The Basic Behaviour
of the Hamilitonian.}

When virtual double occupation of copper sites is excluded so
that only virtual Cu$^+$ excitations are permitted the order $t^2$
Hamiltonian becomes in the limit $U_p \to \inf$,
$$ H =
-t^2\sum_{<ij>} f_{ij}^{-1}(1-P_j) + t^2 \sum_{<i;jj'>\s\s'} \pdp\dd
{\tilde f_{ijj'}^{-1}} d_{i\s'}\p
\eqno(3.1a) $$
$$
f_{ij} = \D + V - V \sum_{<i;jl>}P_l
\eqno(3.1b)
$$
$$
{\tilde f_{ijj'}} = \D  - V \sum_{<i;jj'l>}P_l.
\eqno(3.1c)
$$
The inversion of $f$ leads to the following terms
in the Hamiltonian:
$$
H_0 = -4A_1 \sum_i D_i,
\eqno(3.2a)
$$
$$
H_1 = B_1 \sum_{<ij>} D_i P_j + C_1 \sum_{<i;jj'>\s\s'}
\dd d_{i\s'}\pdp\p,
\eqno(3.2b)
$$
$$
H_2 =  B_2 \sum_{<i;jj'>} D_i P_j P_{j'}
 + C_2 \sum_{<i;jj'k>\s\s'}
\dd d_{i\s'}\pdp P_k\p,
\eqno(3.2c)
$$
$$
H_3 =  B_3 \sum_{<i;jj'k>} D_i P_j P_{j'} P_k
 + C_3 \sum_{<i;jj'kl>\s\s'}
\dd d_{i\s'}\pdp P_k P_l\p,
\eqno(3.2d)
$$
$$
H_4 =  B_4 \sum_{<i;jj'kl>} D_i P_j P_{j'} P_k P_l,
\eqno(3.2e)
$$
where $<i;jj'k>$ denotes that the three oxygen sites $j,j'$ and $k$
are all distinct and neighbour the copper site $i$, and that the sum
runs over all permutations of the indicies. The operators $\dd$ and
$\pd$ create holes predominately on the copper and oxygen sites,
respectively, but with the optimal amount of hybridisation onto
neighbouring sites. The hole number operators are $D_i = \sum_\s
\dd\d$ $(\equiv 1)$ and $P_j = \sum_\s \pd\p$, respectively. Finally,
the energy scales are:
$$
A_1 = { t^2 \over\D + V}, \hskip 0.3 truein A_{n+1} = {V \over \D +V
-nV}A_n,
\eqno(3.3a)
$$
$$
B_1 = {t^2(\D - 3V) \over\D(\D+V)}, \hskip 0.3 truein B_{n+1} =
{V \over \D - nV}B_n = A_{n+1} - (3-n)A_{n+2},
\eqno(3.3b)
$$
$$
C_1 = { t^2 \over\D }, \hskip 0.3 truein C_{n+1} = {V \over \D
-nV}C_n = A_{n+1} + (1+n)A_{n+2}.
\eqno(3.3c)
$$

The structure of the Hamiltonian is shown in equation (3.2)
where it has been decomposed into terms representing the number of
charge carriers involved. There are two parts to each term. First, a
diagonal part, with coefficient $B$, which represents the static
energy of the copper spin and the surrounding oxygen holes. Second,
an off-diagonal part, with coefficient $C$, which represents the
hopping of an oxygen hole from one site to another in the presence of
the copper spin and the relevant number of additional oxygen holes.
The coefficients $B$ and $C$ are given in equation (3.3), and plotted
as a function of $\vd$ in figure (2). The spin independent terms have
absolute magnitude, but the hole motion terms depend on the
relative phase of the hole states.

We consider first the diagonal coefficients. Figure (2) shows
that these terms are negative (attractive) when $\vd
> \thd$. This may be understood as follows: Recall that the operator
$\dd$ creates a hole localised on a copper site which is maximally
hybridised onto the neighbouring oxygen sites. The correction to the
onsite energy of a `copper' hole when there are no neighbouring
oxygen holes is $-4t^2/(\D+V)$. The factor of 4 arises because there
are four empty sites, and the denominator contains a $V$ because
during the virtual occupation of the oxygen site a repulsion from
the neighbouring copper hole is experienced. Next, consider the
correction to the onsite energy  of a `copper' hole when there is
one neighbouring oxygen hole. This is $-3t^2/\D$. Since $U_p$ is
assumed large one of the oxygen sites cannot be hybridised with,
hence the factor 3. However, now the denominator is just $\D$,
because although the `copper' hole experiences a repulsion from the
neighbouring `copper' hole during the virtual occupation of the
oxygen site, it also ${\it avoids}$ the repulsion arising from the
presence of the oxygen hole. By simililar arguments it is easy to
show that the onsite energy corrections are $-2t^2/(\D-V)$ and
$-t^2/(\D-2V)$ for a `copper' hole surrounded by two and three
oxygen holes, respectively. If there are four oxygen holes
surrounding a copper site, then the `copper' hole cannot hybridise
at all.

Now consider the energy difference between having two
oxygen holes neighbouring the {\it same} copper site, and two
seperated oxygen holes. This is:
$$
-{4t^2\over \D+V}-{2t^2\over \D-V} + {6t^2\over\D}
$$
$$
=  {2t^2 V(\D - 3V)\over(\D-V)\D(\D+V)}
\eqno(3.4)
$$
which is precisely $B_2$ (when summed over both permutations) and
is negative when $\vd > \thd$. The essential physics, which will
reappear in other sections, is that during the hole's virtual
vacation of the copper site coulomb repulsion is avoided. There is
more coulomb repulsion avoided if there are more oxygen holes
surrounding the same copper site, hence leading to short range
attractive potentials. Clearly, $\vd > \thd$ is an over estimate on
the strength of $V$ because of the blocking of occupied oxygen sites
in the $U_p \to \inf$ limit.

Turning now to the off-diagonal terms, the matrix element to
hop an oxygen hole from one site to another when there are $n$ other
holes present is $
{t^2\over \D -nV}$, or $0$ if $n=4$.
Thus the amplitude of the matrix elements increases if there are
more holes present. The  reason is the same as above: coulomb
repulsion is avoided during the virtual Cu$^+$ excitation. This is of
course the `bare' matrix element and does not indicate the loss of
kinetic energy resulting from occupied sites.

The question as to whether there are attractive channels for $\vd
\le  \thd$ depends on the sign acquired by the phase of the
off-diagonal terms. This requires diagonalising the Hamiltonian
(3.2) which we address in the following sections.

The physics of the dominant interactions shown in figure (2) are
clear. As the nearest neighbour repulsion is increased there is a
transition from a basically single particle picture with weak two
particle interactions into a regime where three particle
interactions dominate. In $\S 4$ we look at the three particle
regime involving clustering. When the single particle energies
dominate we expect the particles to delocalise and in the low
density limit the two particle interactions will be the second most
important interaction. In $\S 5$ we consider this more realistic
picture.

\head{4. The Clustering Limit.}

In this section we analyse the behaviour of the Hamiltonian (3.2) in
the limit where the three and four particle energy scales dominate.
There is `charge clustering' in this limit. The energy gain from
delocalising the holes is dominated by the short range coulomb
effects, and bound states of several holes occur. We consider the
questions of how many holes cluster together, and whether there are
any residual dynamics.

There are some obvious physical questions to resolve before we
proceed to a solution in this limit. Charge clustering is forbidden
by the long range nature of the coulomb interaction, and so the
effects that we are describing are to a certain extent `unphysical'.
Although long range coulomb interactions are omitted from a tight
binding description, because charge fluctuations are assumed
screened, if this contribution were included it may not invalidate
clustering on short length scales. On long length scales the
charge remains homogeneous, while locally the charge clusters into
either `bubbles' or a sort of `honeycomb' structure. If this
picture is seriously considered then we must ask whether such a
pattern could be dynamic, and hence lead to local charge
fluctuations and a possible exotic type of superconductivity. We do
not take these possibilities seriously, but the wealth of possible
phenomena exhibited by our description deserves discussion.

There are two distinct types of clustering with a transition
between these two types of behaviour when $2V=\D$. One type of
clustering occurs when $2V>\D$ and results in a complete change
of ground state. This is considered in $\S 4.1$. The other type
of clustering occurs when $2V$ approaches $\D$ from below, and is
discussed in $\S 4.2$. The behaviour at the transition, $2V=\D$,
is pathological: the three and four body energy scales diverge and
there appears to be a discontinuous change in hole density.

4.1 Static Charge Clustering: $2V > \D$.

In this limit there is a static change in the ground state provided
that there is a sufficient number of holes. Copper holes vacate their
sites and reside on neighbouring oxygen sites in order to avoid the
local coulomb repulsion, V. These vacated copper sites cluster
together, along with any added oxygen holes, to form a region of the
lattice with singly occupied oxygen sites and vacant copper sites.
Such a configuration obviously avoids the local repulsion, V, except
at any boundary between regions. In a macroscopic cluster there are
two holes per unit cell corresponding to one doped oxygen hole
together with the original copper hole in the undoped phase.
Comparing the energy per added hole in the macroscopic cluster to
that of an isolated hole on an oxygen site immediately gives the
criterion $2V=\D$ for the phase transition. This corresponds to the
balance between the local coulombic repulsion avoided and the loss
from expelling a hole from a copper to an oxygen site.

This argument ignores boundary effects which become important if the
cluster were to split up into smaller droplets. Indeed, a droplet
of six holes is only stable if $V> 3\D$, and a droplet of forty
holes is only stable if $V> 5\D/6$.

Although this limit is not a physically realistic one, it
illustrates the clustering of oxygen holes being driven by the
copper-oxygen repulsion. In this case there is a static change
of state from Cu$^{2+}$ $\to$ Cu$^+$. This transition arises
because coulomb repulsion between holes is  avoided
 ${\it provided}$ that the  holes on oxygen sites cluster
together. This  result is a direct consequence of the unusual
topology of the copper-oxide planes in that copper sites have
four neighbouring oxygen sites, whereas oxygen sites have only two
neighbouring copper sites. For smaller values of $V$  the Cu$^{2+}$
$\to$ Cu$^+$ fluctuations are virtual. Neverthless the same
arguments apply; coulomb repulsion is avoided if the oxygen holes
cluster around the same copper site, implying short range attractive
potentials.

4.2 Dynamic Charge Clustering: $2V<\D$.

As $2V$ aproaches $\D$ from below the three and four body terms
dominate. There is a complication, however, in that the four
particle term is necessarily repulsive (in the large $U_p$ limit)
and so the simple static picture of all the oxygen sites in some
region being filled is invalid. There is a dynamic contribution to
the energy, since the resonant energy is only saved when a copper
site is surrounded by three oxygen holes. The physical source of
this problem is easy to understand; if all four oxygen sites
are occupied the copper hole is unable to move and so there is no
virtual process to resonate.

Now, the static terms of (3.2) are
independent of spin, but the dynamic terms depend on the spin
coherence of the copper and oxygen holes. If we regard the
resonant term as arising from the motion of vacancies in a back
ground of holes then by invoking the Nagaoka theorem\refto{n} we can
determine the spin coherence. Examination of the Hamiltonian (3.2d)
shows that vacancy movement is unfrustrated, with negative
hopping amplitude. The Nagaoka theorem then informs us that the
maximum kinetic energy is gained (in the low density limit) with a
ferromagnetic spin background. Consequently, we restrict our
attention to ferromagnetism, asumming that in a cluster of copper and
oxygen holes all the spins are polarised.

The Hamiltonian (3.2d, 3.2e) then becomes:
$$
H=-A_4\sum_{<i;jj'k>}P_jP_{j'}P_k +3A_4\sum_{<i;jj'kl>} p_j^{\dag}
P_kP_l p_{j'} + A_4\sum_{<i;jklm>}P_jP_kP_lP_m.
\eqno(4.1)
$$
This is not a particularly transparent representation and so we
consider a description based around the cluster limit, where the
number of holes is dense and we may expand away from the case where
each oxygen site has one hole. The excitations are now particle
like and we denote a particle by the creation operator $q_j^{\dag}$.
In terms of the new vacuum the Hamiltonian (4.1) is:
$$
H=-A_4\sum_{<i;jklm>}[Q_J(1-Q_k)(1-Q_l)(1-Q_m) +3q_j^{\dag}
q_k(1-Q_l) (1-Q_m)].
\eqno(4.2)
$$
This Hamiltonian is readily understood. Once the multiple counting
of the terms has been extracted, we find that a single particle can
move to any of the four oxygen sites surrounding a neighbouring
copper site with hybridisation $-6A_4$, provided that it is the
only particle neighbouring that copper site.

The difficulty in solving (4.2) is deciding upon the local density of
holes. If the holes cluster together and form a dense region then
the representation (4.2) becomes the natural description. This
Hamiltonian then describes the motion of vacancies in a cluster
almost full of holes. If there were no vacancies there would be no
contribution, and this is just the realisation of the fact that a
copper hole cannot move if all four neighbouring oxygen sites are
full. In order to gain kinetic energy there must be vacancies, but
how many ? The cluster will assume a size which allows the optimal
gain in vacancy kinetic energy. To determine what this might be we
consider the evidence from different clusters.

We have solved the cluster problems corresponding to the
configurations depicted in figure 3. In order to compare the
results we look at the gain per hole. For the plaquette of figure
3(a) there is one vacancy with an energy gain of $-8A_4$ per hole.
For the two connected plaquettes of figure 3(b) there are two
vacancies with a gain of $-8.4A_4$ per hole. The larger cluster is
relatively stable. The star of five copper atoms (figure 3(c)) has
four vacancies with a gain of $-8.3247A_4$ per hole. Finally, the
square of figure 3(d) has four vacancies with an energy of $-9A_4$
per hole. Obviously the ground state will be a macroscopic cluster.
The density of vacancies is probably not far from a value of half a
hole per cell, and hence we would predict a discontinuity in
the density of holes when $2V=\D$. The loop problems of figure
3(e-h) show the surprising result that the best solution is
independent of the length of the loop at $-9A_4$. We postulate that
this is also the infinite loop result. Considering a square array of
nine copper sites with the eight `edge' sites considered as a loop
and the central site as a `perturbation' should immediately convince
the reader that a more compact cluster is better than a loop.

It should be borne in mind that the gain in energy from extending
the size of the cluster is minor and could easily be overcome,
leading to a delocalised triple of holes. The long range
electrostatic energy prefers homogeneity of charge, so this
contribution could easily be invoked to suggest delocalisation.
Indeed a picture of a homogeneoous distribution of isolated lines
of holes gaining $-9A_4$ energy per hole is an intriguing prospect
of simultaneously ensuring local density fluctuations and global
inhomogeneity of charge. These kinds of phase seperation are
reminiscent of the phase seperation being suggested in the `t-J'
model\refto{em2}, although their origins are quite different.

There are two types of delocalisation in this description. First,
the motion of vacancies, $q_j^{\dag}$, in a dense cluster of holes,
and second the motion of isolated triples and the boundary of any
cluster. The first effect seems very unlikely to lead to a
superconducting instability, and it is only the second effect which
needs to be seriously considered. Fluctuations in the boundary of
any cluster are certainly charged and could, in principle, lead to
superconductivity if there were long range phase coherence which
induces some form of gap. However, we do not believe that these
limits have been realised in the presently known experimental
systems.

\head{5. The Delocalised Limit}

We now consider the limit which is easiest to justify
experimentally. This is to assume that the nearest neighbour coulomb
repulsion is much smaller than the charge transfer gap, $V << \D$.

Although we are predominately interested in the two particle
interactions (3.2c) the `single' particle terms (3.2b) dominate in
this limit. Furthermore, it is these terms (along with the
superexchange term) which determine the underlying spin
correlations. So, before proceeding to discuss the two particle
interactions we
 study one charge carrier first. As has been discussed
elsewhere\refto{l, b}, the motion of a single oxygen hole in the
Cu$^+$ limit is rather subtle.  Below we briefly analyse this in the
$U_p \rightarrow \inf$ limit; a full account of the $U_p=0$ limit is
given in Barford\refto{b}.

The relevant Hamiltonian is (3.2b):
$$ H_1 =
B_1\sum_{<ij>}D_iP_j + C_1\sum_{<i;jj'>\s}\dd d_{i\s'}\pdp\p.
\eqno(5.1)
$$
The
first term  yields a constant energy for each oxygen hole provided
that each copper site is occupied, so it is ignored.
The ground
and excited state are easily found for one plaquette (namely, a
copper site surrounded by four oxygen sites). The ground state is a
singlet spin configuration with the charge delocalised:
$$
|S_i> = {1\over 2
{\sqrt 2}} \sum_{j \in i \s} {\rm sgn}(\s)\dd p^{\dag}_{j {\bar \s}}
|0> \eqno(5.2a)
$$
with delocalisation energy $-3C_1$. This corresponds to the `Zhang
Rice' singlet\refto{zr}. The lowest lying triplet excitation is
triply degenerate:
$$
|T_i> = {1 \over {\sqrt 2}} \dd(\pd - p^{\dag}_{j'\s})|0>
\eqno(5.2b)
$$
for some pair $j,j'$ at the delocalisation energy $-C_1$.

The problems emerge when we try to delocalise the hole on the
lattice. A hole on an oxygen site is a nearest neighbour to two
copper sites. Although it is possible for the oxygen spin to be
simultaneously in a spin triplet with respect to these copper
sites, and hence generate a ferromagnetic state, it is not possible
to simultaneously make the oxygen spin singlet with respect to the
two copper spins.

The purely ferromagnetic solution has the spectra as shown in
equation (5.3), where $2a$ is the copper-copper spacing. In addition
to the bonding band there is a non-bonding band at $-2C_1$:
$$
\epsilon^{\rm non-bonding}_{\underline k} = -2C_1,
\eqno(5.3a)
$$
$$
\epsilon^{\rm bonding}_{\underline k} = -2C_1 + 2C_1\left[ \cos
(2k_x a) + \cos (2k_y a)\right].
\eqno(5.3b)
$$

Evidently, even restricting the local singlet to one plaquette
gives a lower energy than the best ferromagnetic solution. At first
sight we should therefore discard ferromagnetism and study more
reasonable situations. We will not do this, however, because of two
reasons. First, if all the spins are parallel then the copper
spin system becomes passive and we may consider the charged oxygen
system in isolation. This gives a way of seperating the spin
fluctuation induced many body particle interactions from the direct
charge fluctuation induced interactions. Second, by diagonalising
the two body Hamiltonian around a single plaquette we derive the
effective two body potentials. We look at the two particle
interactions in the ferromagnetic state in $\S 5.1$, and consider
the low spin scenario in $\S 5.2$, summarising in $\S5.3$.

5.1 Ferromagnetism

The two particle interactions are now simply,
$$
H_2 = B_2\sum_{<i;jj'>}P_jP_{j'} + C_2\sum_{<i;jj'k>}
p_{j'}^{\dag}P_k p_j.
\eqno(5.4)
$$

In the limit of interest,
$A_1 > A_2 >A_3$, the first term is a repulsion between pairs of
holes on the same plaquette. The question arises: is there a two
particle wavefunction for which the second term is attractive and
dominates the first term?

In the diagonal representation the two particle Hamilitonian is:
$$
H_2 = 4A_2 \sum_i\sum_{\mu=1}^3 b_{\mu}^{i\dag}b_{\mu}^i -
8A_3\sum_i\sum_{\mu=1}^3
 c_{\mu}^{i\dag}c_{\mu}^i,
\eqno(5.5)
$$
where
$$
 b_{\mu}^{i\dag} = {1 \over 2}(p_{\alpha}^{i\dag} +
p_{\beta}^{i\dag} )(p_{\gamma}^{i\dag} + p_{\delta}^{i\dag}),
\eqno(5.6a)
$$
$$
 c_{\mu}^{i\dag} = {1 \over 2}(p_{\alpha}^{i\dag} -
p_{\beta}^{i\dag} )(p_{\gamma}^{i\dag} - p_{\delta}^{i\dag}),
\eqno(5.6b)
$$
the sum $i$ is over plaquettes and the indicies $\{
\alpha,\beta,\gamma,\delta \}$ label the different oxygen sites
around the $i$th. plaquette. The index $\mu$ corresponds to the
three  ways that the four oxygen sites in the plaquette can split
into two pairs, $\alpha,\beta$ and $\gamma, \delta$.

Evidently, in a system with purely repulsive interactions we have
found an attractive pairing. Pairs of holes in the $c_{\mu}^{\dag}$
configuration feel an attraction with an eigenvalue of $-8A_3$,
which corresponds to a coupling of $\sim 0.5$ eV, using the data in
$\S 1$.

The dominant effect is a repulsion between a pair made up of a hole
in phase on one pair of sites and another hole in phase on the
other pair of sites. The weaker attractive combination is the
anti-phase combination of the two holes. If we consider
$c_{\mu}^{\dag}$ as a cooper pair then we may deduce something about
the symmetry of the superconducting gap function.

Let us now reconstitute the lattice. If we consider the two particle
interaction in isolation, then the natural cooper pair has one
member on one oxygen sublattice and the other member on the other
sublattice. Since the pairs have parallel spins the spin symmetry
is symmetric. The sublattice degree of freedom is anti-symmetric
and so the final spatial symmetry is therefore symmetric.  Three
ways of assigning sublattice indicies are shown in figure 4.  Pairs
moving parallel to the cartesian axes would favour the situation
depicted in figure 4(a). Holes with a uniform wavefunction would
feel the repulsion mediated by $b_{\mu}^{\dag}$, whereas holes with
alternating phase would feel the attraction mediated by
$c_{\mu}^{\dag}$.  A similiar situation holds in figure 4(b). Figure
4(c) shows holes on the `squares' interacting with each other. The
relevant situation to consider is that corresponding to holes at the
Fermi surface, for these can make use of the weak attractive
interaction at little loss in kinetic energy. However, there is a
complication as the lowest band in the ferromagnetic bandstructure
is flat, and so there is an implied degeneracy. The kinetic energy
is optimised using the non-bonding band. Inspection of the
non-bonding solutions resolves the predicament. We observe that a
single hole delocalised on one of the `squares' in figure 4(c) with
an alternating phase around the `square' is an eigenstate of the
Hamiltonian, and corresponds to a real space non-bonding orbital.
The alternating phase is perfect for gaining the two body attraction
and so we find that the ground state for the ferromagnet, with weak
two body interactions included, is composed of localised non-bonding
states clustered together in a pattern similar to figure 4(c).
However, this is still not a scenario for superconductivity as the
states are localised.

There are bonding states which are degenerate with the non-bonding
states; those for which $\gamma_{\underline k}= \pm(\pi/2a)(1,1)$.
These states have precisely the alternating symmetry of figure 4(b)
and so could conceivably carry current.
A close inspection of the wavefunction shows that `s-wave' pairs
suffer the repulsion while the `d-wave' pairs feel the weak
attraction.

5.2 Low Spin

For three spins the low spin scenario is total spin 1/2. We analyse
this case to try to deduce the sort of
cooper pair to be expected in a low spin ground state.

The `two' particle interactions are:
$$
H_2 = B_2\sum_{<i;jj'>}D_iP_jP_{j'} + C_2\sum_{<i;jj'k>}
\dd d_{i\s'} \pdp P_k \p.
\eqno(5.7)
$$
In a total spin 1/2 subspace the diagonalised Hamiltonian is:
$$
H_2=4A_4\sum_i\sum_{\mu=1}^4 e_{\mu}^{i\dag}e_{\mu}^i+
(3A_2-2A_3)\sum_i\sum_{\mu=1}^3 f_{\mu}^{i\dag}f_{\mu}^i+
$$
$$
(A_2-6A_3)\sum_i\sum_{\mu=1}^2 g_{\mu}^{i\dag}g_{\mu}^i-
(A_2-10A_3)\sum_i\sum_{\mu=1}^3 h_{\mu}^{i\dag}h_{\mu}^i.
\eqno(5.8)
$$
Again we find an attractive channel, now corresponding to the $
h_{\mu}^{i\dag}$ combination.
$$
h_{\mu}^{i\dag}={1\over {\sqrt 10}}\left[
(p_{\alpha \uparrow}^{i\dag}+
p_{\beta \uparrow}^{i\dag})(p_{\gamma \uparrow}^{i\dag}+
p_{\delta \uparrow}^{i\dag})d_{i \downarrow}^{\dag}-
{1\over 2}\sum_\s (p_{\alpha \s}^{i\dag}+p_{\beta \s}^{i\dag})
(p_{\gamma {\bar \s}}^{i\dag}+p_{\delta {\bar \s}}^{i\dag})
d_{i \uparrow}^{\dag}\right]
$$
$$
+ {1 \over {\sqrt 5}}\left[
 {1 \over {\sqrt 2}} \sum_\s {\rm sgn}(\s)
p_{\gamma \s}^{i\dag}p_{\delta {\bar \s}}^{i\dag}-
{1 \over {\sqrt 2}} \sum_\s {\rm sgn}(\s) p^{i\dag}_{\alpha \s}
p_{\beta {\bar \s}}^{i\dag}\right]  d_{i \uparrow}^{\dag},
\eqno(5.9a)
$$
and, as before, $\{ \alpha,\beta,\gamma,\delta \}$ labels the
oxygen sites, and the $\mu$ index runs over the possible ways of
splitting the sites into pairs. There is another representation for
this quantity which is not orthogonal but which is very suggestive:
$$ h_{\mu}^{i\dag}={1\over 2 {\sqrt 5}}\left[(p_{\alpha
\uparrow}^{i\dag}+ p_{\beta \uparrow}^{i\dag})
{1 \over {\sqrt 2}} \sum_\s {\rm sgn}(\s) (p_{\gamma \s}^{i\dag}+
p_{\delta \s}^{i\dag})d_{i {\bar \s}}^{\dag}-
(p_{\gamma \uparrow}^{i\dag}+
p_{\alpha \uparrow}^{i\dag})
{1 \over {\sqrt 2}} \sum_\s {\rm sgn}(\s) (p_{\alpha \s}^{i\dag}+
p_{\beta \s}^{i\dag})d_{i {\bar \s}}^{\dag}\right]
$$
$$
+{1 \over {\sqrt 5}}\big[ p_{\alpha \uparrow}^{i\dag}
{1 \over {\sqrt 2}} \sum_\s {\rm sgn}(\s) p_{\beta \s}^{i\dag}
d_{i {\bar \s}}^{\dag}+p_{\beta \uparrow}^{i\dag}
{1 \over {\sqrt 2}} \sum_\s {\rm sgn}(\s) p_{\alpha \s}^{i\dag}
d_{i {\bar \s}}^{\dag}
$$
$$
- p_{\gamma \uparrow}^{i\dag}
{1 \over {\sqrt 2}} \sum_\s {\rm sgn}(\s) p_{\delta \s}^{i\dag}
d_{i {\bar \s}}^{\dag}-
p_{\delta \uparrow}^{i\dag}
{1 \over {\sqrt 2}} \sum_\s {\rm sgn}(\s) p_{\gamma \s}^{i\dag}
d_{i {\bar \s}}^{\dag} \big] .
\eqno(5.9b)
$$
{}From this representation it is clear that the local singlet pairs
of one copper and one oxygen hole are all in phase, while the
`$\uparrow$' contributions are in opposite phases on the two
`sublattices'.

Let us now consider the symmetry of the `cooper pairs'. The object
$h_{\mu}^{\dag}$ does not create a pair of particle, but a triple.
The important point, however, is that only two of these particles,
the oxygen holes, display a charge degree of freedom. The copper
hole displays only a spin degree of freedom and so, although it
participates in the spin symmetry of the cooper pair, it does not
stop the object being doubly charged, and does not participate
directly in the motion of the pair. Considering the two oxygen
holes to be the pair and the copper hole to be a `sink' for
`unwanted quantum numbers' we find that the pair has mixed
character. The natural seperation when the lattice is reconstituted
is again into the two sublattices. We find, from (5.9a), that
two fifths of the wavefunction has the pair in a relative singlet
with symmetric sublattice coupling and symmetric spatial symmetry,
while the other three fifths has the pair in a relative triplet
with antisymmetric sublattice coupling and symmetric spatial
symmetry. This  predicts an `s-wave' spatial gap function.

The final issue to deal with concerns the relation between the
single particle solutions and the pairing interactions. Are the
holes at the Fermi surface of the correct symmetry to make use of
this attraction ? For the ferromagnetic case this was easy; an
analysis of the low lying single particle wavefunction showed that
the local symmetry was precisely the combinations making up the
cooper pairs. We do not have the exact single particle
wavefunctions for the low spin case, so approximate arguments are
necessary.

In order to determine the local correlations in more detail we
performed a ground state calculation for two oxygen holes on the
cluster of figure 5(a) using the `single' particle Hamiltonian
(5.1). The ground state is a total spin singlet, with the four states
pictured in figure 5(b) forming a basis. It is found that the ground
state is predominately $|1>$ and $|2>$ with a small admixture of
$|3>$. The eigenvalue is $-5.6613C_1$ to be compared with the
unfrustrated bound of $-6.0C_1$. The important comparison is between
$|2>$ and $|3>$, which have both holes neighbouring the copper site,
and the state described by $h_{\mu}^{\dag}$.
There is no ${\it a}$ ${\it priori}$ reason why the present
wavefunction should be related to $h_{\mu}^{\dag}$, but there is a
strong resemblance. The phases are such that all the singlets of
copper and oxygen holes are in phase, and the hole on the central
oxygen site in $|2>$ is in anti-phase with the non-singlet
contributions in $|3>$. This is the closest wavefunction to
$h_{\mu}^{\dag}$ which can be made with the restricted symmetry. The
dominant central hole is anti-phase with the two elements of the
other sublattice, in agreement with (5.9b).

5.3 Discussion.

Let us now consider whether or not the attractive
channels found in the above discussions are relevant to the full
many body problem. Since we cannot solve this problem exactly the
discussion will necessarily be rather general.

In the weak coupling
limit the Hamiltonian (3.2) is
$$
H = H_1 + H_2
\eqno(5.10)
$$
with $H_1$ given by (5.1) and $H_2$ by (5.7). Now, $H_1$ is the
dominant term, and implicitly contains many body interactions which
cannot be solved. $H_2$, however, can be solved and contains
explicit interactions. The question is, are these interactions
sympathetic to those of $H_1$ ? To analyse this let us write
$H_2$ as,
$$
H_2 = \sum_i H_2^{(i)},
\eqno(5.11)
$$
where the sum is over all plaquettes. $ H_2^{(i)}$ can be readily
diagonalised, as shown in $\S\S 5.1$ and $5.2$. We therefore
write
$$
H = H_1 + \sum_i  \sum_{\mu =1}^{18} \eps_{\mu} X_{\mu}^{i\dag}
 X_{\mu}^i
\eqno(5.12)
$$
where $\eps_{\mu}$ is the eigenvalue which corresponds to the
eigenfunction $|X^i_{\mu}> = X_{\mu}^{i\dag}|0>$. The operator
$X_{\mu}^{i\dag}$ creates the `cooper pairs' of equations (5.6)
and (5.9). There are four negative eigenvalues (attractive channels)
and fourteen positive ones (repulsive channels). These may be
regarded as being short range attractive and repulsive
two body potentials.

Since $H_1$ is the dominant
term we need to consider which symmetry (two oxygen hole
correlations about the same plaquette) this term prefers. If its
groundstate wavefunction has correlations described by a $|X_{\mu}>$
with a negative eigenvalue, then $H_2$ automatically acts as an
attractive potential, and vice versa.
For the ferromagnetic case we showed that the wavefunction
associated with $H_1$ had precisely the right correlations to
experience the attractive potential. This was confirmed for the low
spin case by the calculation on the cluster of two neighbouring
plaquettes.

The full many body eigenstate of (5.12) will, in general, contain
superpositions of all the basis states which span the Hilbert
space. These basis states will, in turn, have as components the
$|X_{\mu}>$s. From the previous discussion, we expect these terms
to be dominated by the attractive channels. Although we cannot
state the precise symmetry of the cooper pairs, as they will be a
superposition of the $X^{\dag}_{\mu}$s, the cluster calculation
suggests that if the total spin of the complete many body
wavefunction is zero then the cooper pair will be spatially
symmetric. In the high temperature superconductors the motion of
the holes coupled to the Heisenberg term has a dominant effect,
destroying long range magnetic correlations and guaranteeing that
the total spin is in a total spin singlet. This calculation
therefore lends evidence to the suggestion that the charge
fluctuations coupled to the coulomb repulsion result in `s-wave'
pairing.

\head {6. Conclusions.}

The copper-oxide systems are characterised by charge transfer
resonances due to the proximity of the atomic copper and oxygen
energy levels. These charge fluctuations, coupled to the nearest
neighbour copper-oxygen repulsion, $V$, have been investigated by
deriving effective Hamiltonians which describe the low energy
behaviour, starting from the generic two band model. We considered
hole doping only.

When Cu$^{2+}$ $\to$ Cu$^{3+}$ fluctuations dominate we find a
description in which the oxygen hole forms a singlet with the
copper spin. This entity propogates through the square lattice like a
vacancy in a single band Hubbard model. This is equivalent to the
description of Zhang\refto{z}. The effect of the copper-oxygen
repulsion is to cause a nearest neighbour repulsion between these
singlets. The reason for this repulsion is that during the oxygen
hole's virtual occupation of the copper site it feels a repulsion
from all the oxygen holes neighbouring that copper site.

If Cu$^{2+}$ $\to$ Cu$^{+}$ fluctuations dominate, however, we have
shown that short range coulomb repulsion can give rise to attractive
interactions between some of the charge carriers in the system.
This is because during the `copper' hole's virtual vacation of
its site and occupation of a neighbouring oxygen site it avoids
the repulsion from all the oxygen holes neighbouring that copper
site.

The  overall
behaviour is governed by the ratio of the copper-oxygen repulsion to
the charge transfer gap, $\vd$. When $0 \ll \vd \leq {1 \over 2}$,
we proved that charge clustering takes place by showing that
clusters of charge are more stable than isolated charges. On the
other hand, when $0 \leq \vd \ll {1 \over 2}$, the single particle
terms dominate and the question of whether weak coupling
superconductivity exists is a subtle one which we have not been able
to address unambiguously. As we discussed in detail in $\S 5.3$, the
Hamiltonian contains both attractive and repulsive short range two
body potentials. The holes at the fermi surface will only experience
the attractive potentials if they have correlations consistent with
them, otherwise the repulsive interactions will act. On a small
cluster we showed that the correlations are consistent with the
attractive potentials, and indicate `s' wave pairing.
Experimentally, $\vd$ would appear to be about 1/4, placing the
copper-oxide systems in the weak coupling limit.

Finally, let us recall the approximations made in these calculations.
We derived an effective Hamiltonian of $O(t^2)$, thereby neglecting
spin fluctuations. The copper and oxygen coulomb repulsions were
assumed very large, which forbids double occupation of copper
and oxygen sites. The latter constraint is probably too severe, and
tends to work against the attractive mechanisms operative here.
Lastly, the direct oxygen-oxygen hybridisation was neglected.

\vfill\eject

\subhead{References}

\refis{hi} Hirsch J E, 1987, Phys. Rev. Lett., 58, 228.

\refis{ba} Balseiro C A, Rojo A G, Gagliano E R, Alascio B, 1988,
Phys. Rev. B, 38, 9315.

\refis{varma} Varma C M, Schmitt-Rink S and Abrahams E, 1987, Solid
State Comm., 62, 681.

\refis{little} Littlewood P B, Varma C M, Schmitt-Rink S, Abrahams
E, 1989, Phys. Rev. B, 39, 12371.

\refis{grilli} Grilli M, Raimondi R, Castellani C, Di Castro C,
Kotliar G, 1991, Int. J. Mod. Phys. B, 5, 309.  1991, Phys. Rev.
Lett., 67, 259.

\refis{b} Barford W, 1990, J. Phys.: Condens. Matter, 2,  2965.

\refis{lb1} Long M W and Barford W, Rutherford Appleton Laboratory
Reports: RAL-89-053/4.

\refis{l} Long M W, 1988, J. Phys. C, 21 L939.

\refis{lb} Long M W and Barford W, 1989, Physica C, 162-164, 789.

\refis{sw} For a review of the spectroscopic data see, Sawatzky G A,
in ``High Temperature Superconductivity'', ed. Tunstall D P and
Barford W, Adam Hilger, Bristol (1992).

\refis{em2} Emery V J, Kivelson S, Lin H Q, 1990, Phys. Rev. Lett.,
64, 475.

\refis{n} Nagaoka Y, 1966, Phys. Rev., 147, 392.

\refis{zr} Zhang F C and Rice T M, 1988, Phys. Rev. B, 37, 3579.

\refis{em}  Emery V J, 1987, Phys. Rev. Lett., 58, 2794.

\refis{zr} Zhang F C and Rice T M, 1988, Phys. Rev. B, 37, 3579.

\refis{sw} Sawatzky G A, in ``High Temperature Superconductivity'',
ed. Tunstall D P and Barford W, Adam Hilger, Bristol (1992).

\refis{mms} McMahan A K, Martin R M and  Satpathy S, 1988, Phys.
Rev. B, 38, 6650.

\refis{k} Schrieffer J R and Woolf P A, 1966, Phys. Rev. B, 149,
491.

\refis{ck} Kittel C, ``Quantum Theory of Solids'', J Wiley and Sons -
New York (1963).

\refis{fs} Fedro A J and Sch\"uttler H-B, 1989, Phys. Rev. B, 40,
5155.

\refis{z}  Zhang F C, 1988, Phys. Rev. B, 37, 3759.

\refis{n}  Nagaoka Y, 1966, Phys. Rev., 147, 392.

\endreferences

\vfill\eject

\head{Appendix}

The analysis of $\S2$ is completed by inverting the
operator $f_{ij\s}$ of equation (2.1c). The solution may be deduced
from the more general problem:
$$
F = \left[ A - a_i \sum_{i=1}^N P_i \right]^{-1}
\eqno(A1)
$$
where the $P_i$ are mutually commuting projection operators:
$$
P_i^2 = P_i.
\eqno(A2)
$$
Obviously $\dd\d$ and $\pd\p$ are projection operators, but if
double occupancy is prohibited then $D_i = \sum_\s \dd\d$ is
effectively a projection operator since,
$$
D_i^2  - D_i = 2\dd\ddm\dm\d,
\eqno(A3)
$$
and the right hand side only contributes when a site is doubly
occupied.

We may invert the operators iteratively by observing that:
$$
F= (1-P_1)\left[ A - a_i\sum_{i=2}^N P_i \right]^{-1} + P_1\left[ A
-a_1 -a_i \sum_{i=2}^N P_i \right]^{-1},
\eqno(A4)
$$
and hence:
$$
F = \sum_I (A- \sum_{i\in I} a_i)^{-1} \prod_{i \in I} P_i \prod_{j
\ni I}(1-P_j) \eqno(A5)
$$
where the I denotes all the subsets of $\{1,2,...,N\}$.

A final result which is of use in deriving the forms presented in
$\S4$ is,
$$
\left[A- V\sum_{i=1}^N P_i \right]^{-1} = A_0 +A_1\sum_{i_1=1}^N
P_{i_1} +A_2 \sum_{i_1 i_2} P_{I_1} P_{i_2} +
A_3 \sum_{i_1 i_2 i_3} P_{i_1} P_{i_2} P_{i_3} ...
\eqno(A6)
$$
where
$$
A_n = {V \over (A-nV)}A_{n-1}, \hskip 0.5 truein A_0 = {1 \over A}
\eqno(A7)
$$
and $\{i_1 i_2 i_3 \}$ denotes mutually exclusive labels including
all permutations.

\vfill\eject

\subhead{Figure Captions}

Figure 1:

The Cu$^{3+}$ limit with one mobile hole.

(a, b) The natural definitions for local singlets.

(c) The non-bonding combinations of local singlets.

(d) Various non-bonding eigenstates of the Cu$^{3+}$ limit.

(e) The bonding combination of local singlets.

Figure 2:

The relevant energy scales for the induced many body interactions
in this desciption. The subscript denotes the number of oxygen
holes taking part in the interaction. The $B$ and $C$ denote the
coefficients to the diagonal and off-diagonal terms, respectively.

Figure 3:

The configurations of atoms used to investigate the clustering
limit. The `loops' in figures (e-h) denote bonds forming periodic
boundary conditions.

Figure 4:

Several ways of breaking the oxygen atoms up into sublattices.

(a) Two sublattices with pure square lattice symmetry.

(b) Two sublattices which break the square lattice symmetry. Using
standing waves composed from $\pm {\pi \over 2a}(1,1)$, which
reside on the non-interacting Fermi surface, we can isolate
particles on one or other of the two sublattices.

(c) Two sublattices where the `squares' relevant for non-bonding
orbitals play the dominant r\^ole.

Figure 5:

(a) The configuration of atoms chosen to investigate the two oxygen
hole wavefunction in the absence of nearest neighbour repulsion.

(b) A pictorial representation of the symmetries of the states
which form the basis described in the text.

\end